\documentclass[twocolumn,aps,prl,longbibliography]{revtex4-1}

\usepackage{amssymb}
\usepackage{amsmath,latexsym}
\usepackage{graphicx}
\usepackage{dsfont}
\usepackage{color,soul}
\usepackage{multirow}

\usepackage[%
colorlinks=true,
urlcolor=blue,
linkcolor=blue,
citecolor=blue
]{hyperref}

\begin{document}
	
\title{Stepped Graphene-based Aharonov-Bohm Interferometers}
	
\author{V. Hung Nguyen\footnote{E-mail: viet-hung.nguyen@uclouvain.be} and J.-C. Charlier} \address{Institute of Condensed Matter and Nanosciences, Universit\'{e} catholique de Louvain, Chemin des \'{e}toiles 8, B-1348 Louvain-la-Neuve, Belgium}
	
\begin{abstract}
	Aharonov-Bohm interferences in the quantum Hall regime are observed when electrons are transmitted between two edge channels. Such a phenomenon has been realized in 2D systems such as quantum point contacts, anti-dots and \textit{p-n} junctions. Based on a theoretical investigation of the magnetotransport in stepped graphene, a new kind of Aharonov-Bohm interferometers is proposed herewith. Indeed, when a strong magnetic field is applied in a proper direction, oppositely propagating edge states can be achieved in both terrace and facet zones of the step, leading to the interedge scatterings and hence strong Aharonov-Bohm oscillations in the conductance in the quantum Hall regime. Taking place in the unipolar regime, this interference is also predicted in stepped systems of other 2D layered materials.
\end{abstract}
	
\maketitle
	
The fascinating properties of quantum hall devices arise from their ideal 1D edge states formed in a 2D electron system when a high magnetic field is applied \cite{Klaus1986,Ferry2015}. These edge states are particularly attractive due to their large coherence lengths, which is mandatory for constructing electron interferometers. However, since the edge channels are spatially separated, a mechanism for creating the electron transmission between them is required to achieve the interference effects. In this regard, one explored technique consists in building  constrictions (quantum point contacts) in a sample, where the interedge tunneling paths can occur \cite{Alphenaar1992,Chamon1997,Halperin2011,Bonderson2006a,Bonderson2006b,Stern2006,Ilan2009,Camino2007a,Camino2007b,Zhang2009,McClure2009,Lin2009,Ofek2010,McClure2012,Sabo2017}. Setups consisting of a pair of quantum point contacts with an internal cavity has been demonstrated to work well as quantum Hall, electronic Fabry-P\'{e}rot, and Aharonov-Bohm interferometers. Another mechanism has also been suggested in systems consisting of an antidot introduced between their edges \cite{Kataoka2000,Sim2003,Ihnatsenka2009,Sim2008,Hackens2010,Paradiso2012,Martins2013,Sarah2016}. Electronic currents encircling the antidot can be achieved and a similar Aharonov-Bohm (AB) interference is hence observed.
	
Graphene, a truly 2D material, is an ideal platform for investigating quantum Hall and interference effects. Remarkably, owning to an unique linear dispersion and Dirac-like fermions \cite{Neto2009}, Landau levels and a half-integer quantum Hall effect with an unusual quantized sequence compared to the conventional systems have been observed in graphene when a strong magnetic field (\textit{B-}field) is applied \cite{Yin2017,Machida2015}. In addition, with its semimetal character, quantum Hall systems of graphene can work in both the unipolar and bipolar regimes that can be generated and controlled by gate voltages \cite{Machida2015,Abanin2007,Williams2007,Ozyilmaz2007,Morikawa2015,Kolasinska2016,Wei2017,makk2018}. Interestingly, in the bipolar regime the chiral edge states equilibration and interedge scatterings at the \textit{p-n} interfaces in graphene have been observed, resulting in fractional conductance plateaus \cite{Ozyilmaz2007}.
	
With its typically high carrier mobilities, graphene is also an ideal material to perform the investigation on interference effects, including the Aharonov-Bohm one. Several experimental and theoretical observations of the AB effect in graphene nanorings have been reported (i.e., see Ref. \cite{Schelter2012} and references therein). Remarkably, graphene \textit{p-n} junctions can also work as AB interferometers \cite{Morikawa2015,Kolasinska2016,Wei2017,makk2018} in the quantum Hall regime. In particular, the oppositely propagating edge states are formed in two different doped zones and their interaction at the \textit{p-n} interface acquires conductance oscillations of the AB periodicity at high \textit{B-}fields.
\begin{figure}[!b]
	\centering
	\includegraphics[width = 0.48\textwidth]{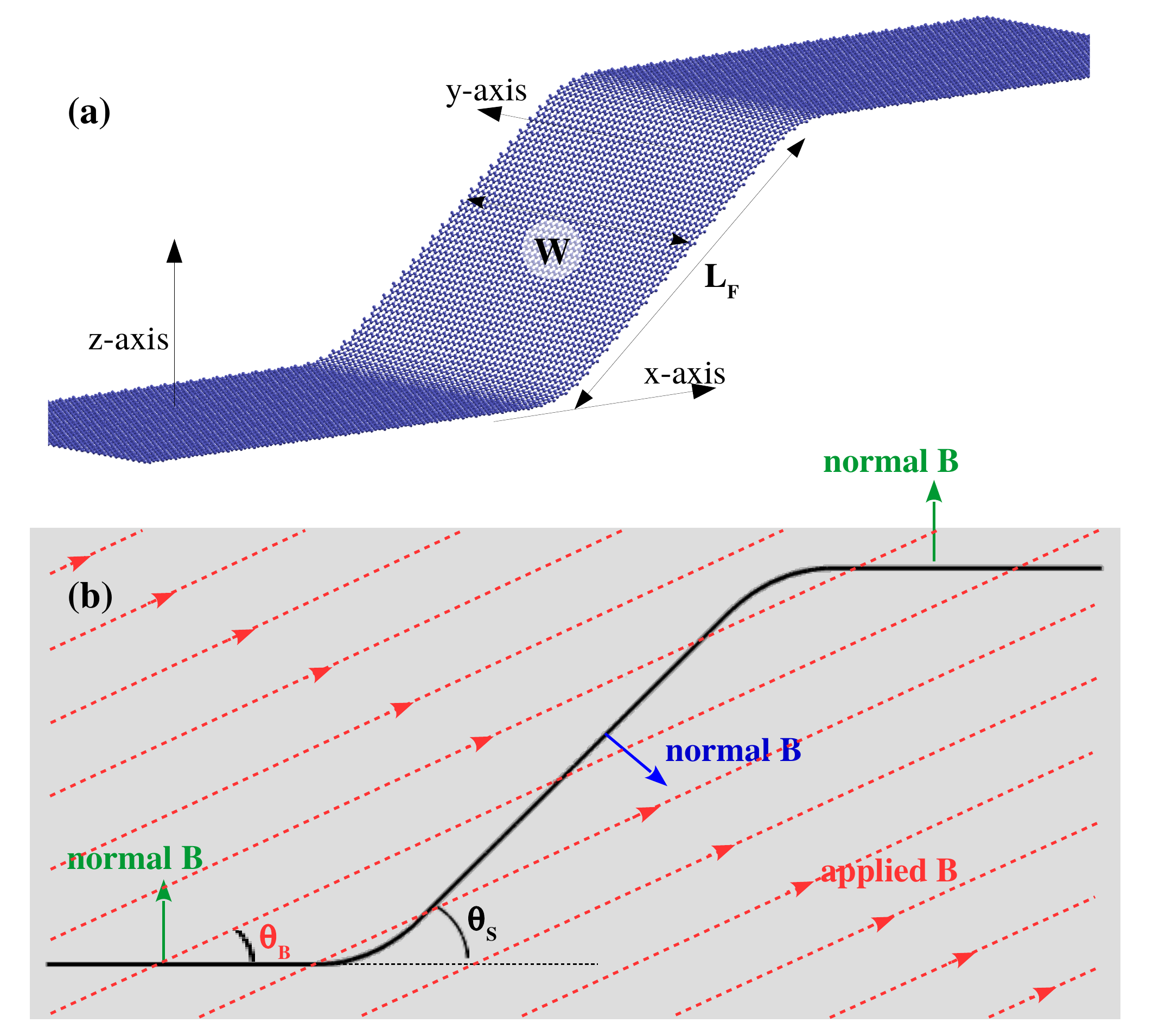}
	\caption{Stepped graphene (a) with the ribbon width $W$ and the length of facet zone $L_F$. (b) Schematic of the side view illustrating the applied magnetic field (red dashed lines of arrows) and its normal components (green and blue arrows, respectively) in both terrace and facet zones. $\theta_B$ and $\theta_S$ are the angles of the field and of the facet zone relative to the terrace one (Ox axis), respectively.}
	\label{fig_sim1}
\end{figure}
	
Motivated by such scientific context, a new kind of Aharonov-Bohm interferometers based on stepped graphene channels is proposed herewith. These non-planar systems have been actually achieved in several experimental situations. For example, the step bunching on the SiC surface is often observed in epitaxial graphene growth by thermal decomposition of SiC \cite{Odaka2010,Kuramochi2012,Schumann2012,Pallecchi2014,Ciuk2014,Akira2015,Giannazzo2014,Bao2016,Speck2017,Stohr2017,Pakdehi2018}, a promising method for the production of large-area high-quality graphene. These stepped graphene channels with terrace size of several $\mu m$ and step height of tens \textit{nm} can be controllably produced by varying the heating rate \cite{Bao2016}. Non-planar graphene systems have been also synthesized in an even better controllable way by draping graphene on pre-structured substrates \cite{Briggs2010,Takahiro2012,Kenjiro2012,Lee2013,Lee2014,Ayhan2014,Dai2015,Bai2017}.

In this work, a $B$-field is found to induce different effects on the electron motion in the terrace and facet zones of stepped graphene, essentially resulting from the creation of different normal components of the field in these zones. This feature has been also observed in several non-planar systems of graphene  \cite{Ciuk2014,Odaka2010,Kuramochi2012,Schumann2012,Pallecchi2014,Akira2015,Pakdehi2018} and 2DEG \cite{Leadbeater1995,Leadbeater1996,Ibrahim1997,Cina1999,Nauen2002,Grayson2004}, leading to anisotropic pictures when the transport takes place in the directions aligned parallel and perpendicular to the step edge. Here, a novel phenomenon is predicted when tuning the direction of $B$-field applied to stepped graphene. In particular, an inhomogeneous profile containing alternatively opposite normal \textit{B-}components along the channel can be created, thus inducing accordingly opposite edge states. The interaction between these edge states finally results in strong AB oscillations in the quantum Hall regime as presented in this article.

The considered systems consist in graphene nanoribbons (GNRs) in the step geometry as illustrated in Fig.\ref{fig_sim1}. In general, in-plane local strains can occur in the bent zones of the step, however, have been demonstrated to be small (i.e., $< 1\%$) and negligible even in epitaxial graphene \cite{Robinson2019,Briggs2010}. Moreover, such small local strains are shown not to strongly affect the predicted AB interference picture \cite{supplmater}. Therefore, these local strains are neglected and the $p_z$ tight-binding Hamiltonian \cite{Neto2009} is employed to compute the magneto-transport in this work. In particular, when a \textit{B-}field is applied,
\begin{equation}
	H =  \sum\limits_{n}  U_n c^\dagger_n c_n + t_0 \sum\limits_{\left\langle n,m \right\rangle} e^{i\phi_{nm}} c^\dagger_n c_m
\end{equation}
where $U_n$ represents the potential energy at the n$^{th}$ site, $t_0 = -2.7$ eV corresponds the nearest-neighbor hopping energies, and $\phi_{nm} = \frac{e}{\hbar} \int_{\mathbf{r}_n}^{\mathbf{r}_m} \mathbf{A}(\mathbf{r}) d\mathbf{r}$ is the Peierls phase describing the effects of the \textit{B}-field. Here, the magnetic field $\mathbf{B} = B (\cos\theta_B,0,\sin\theta_B)$ is considered by introducing the vector potential $\mathbf{A}(\mathbf{r}) = -B(y\sin\theta_B,z\cos\theta_B,0)$. The above Hamiltonian is solved using the Green's-function technique \cite{Nguyen2013a,Lewenkopf2013,supplmater}, allowing for the calculation of the transport quantities perpendicular to the step edges, i.e., along the Ox axis shown in Fig.\ref{fig_sim1}.
	
\begin{figure}[!b]
	\centering
	\includegraphics[width = 0.49\textwidth]{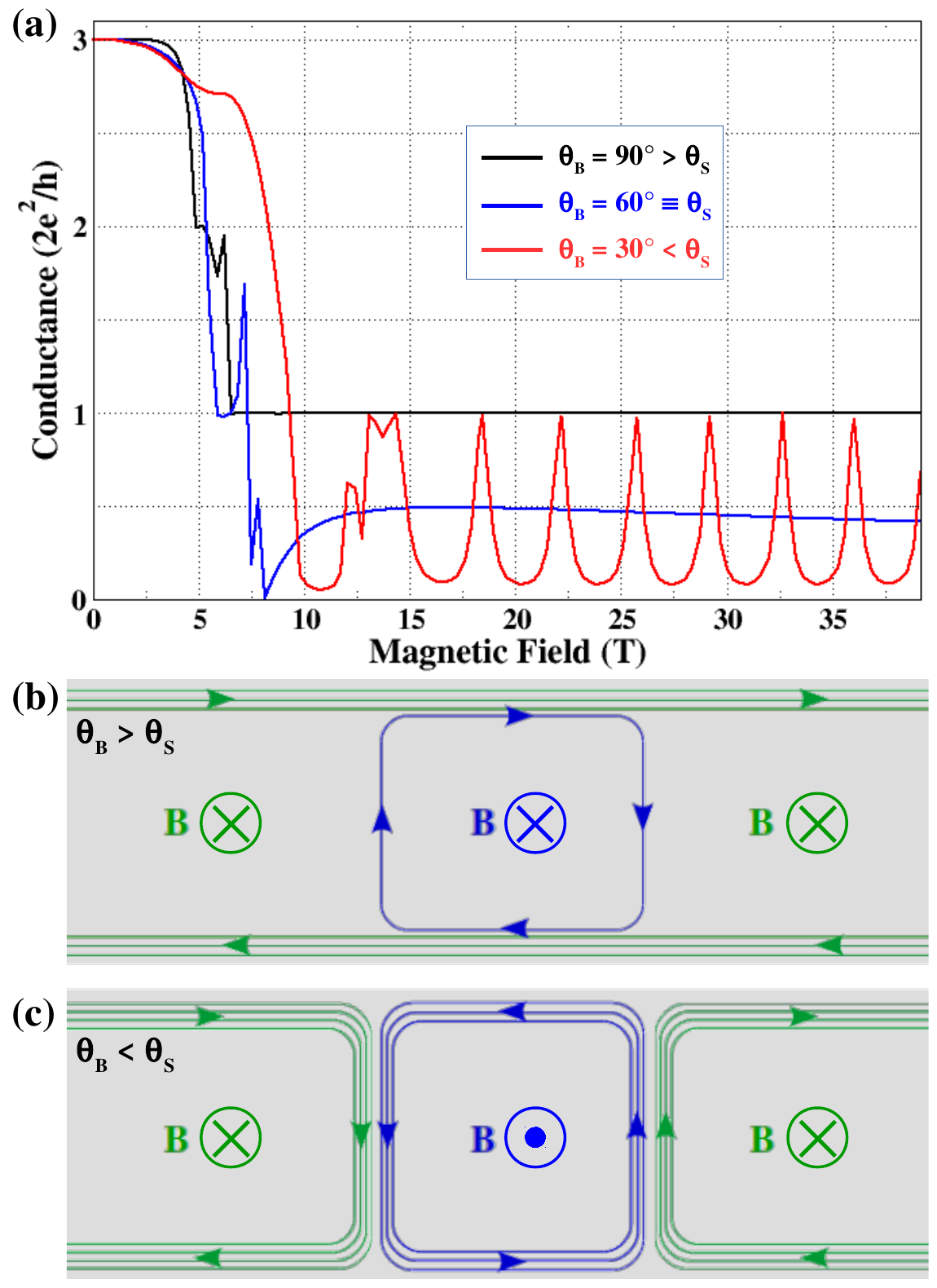}
	\caption{(a) Conductance as a function of $B$-field in a stepped armchair GNR for different angles $\theta_B$ and with $\theta_S = 60^\circ$, $E_F = 75$ $meV$, $L_F \simeq 75$ $nm$, and $W \simeq$ 40 $nm$ (i.e., number of dimer lines $N_a = 324$, a semiconducting GNR). (b,c) Diagrams illustrating the interedge scatterings for $\theta_B > \theta_S$ and $\theta_B < \theta_S$, when the normal \textit{B-}components in the facet and terrace zones (see Fig.\ref{fig_sim1}b) are pointing out in the same and opposite directions, respectively.}
	\label{fig_sim2}
\end{figure}

Fig.\ref{fig_sim2}a displays the dependence of conductance on $B$-field applied in different directions in a step constituted by an armchair GNR. As mentioned, the applied $B$-field induces different normal components ($B_N$) in the terrace and facet zones, i.e., $B_N = B\sin\theta_B$ (green arrows) and $-B\sin(\theta_S-\theta_B)$ (blue arrow), respectively (see Fig.\ref{fig_sim1}b). First, for $\theta_B > \theta_S$, $B_N$-components pointing out in the same direction are obtained, thus inducing the same propagating edge states in the two zones as illustrated in Fig.\ref{fig_sim2}b. Therefore, when a large $B$-field is applied, a conventional Landau quantization is still obtained, i.e., the conductance represents quantized values as in Fig.\ref{fig_sim2}a for $\theta_B = 90^\circ$. For $\theta_B \equiv \theta_S$, the $B_N$-component in the facet zone is canceled and hence the $B$-field has no effect on the in-plane transport in this zone. Consequently, the system behaves as a heterojunction consisting of finite- and zero-magnetic field zones and the scatterings at their interface basically explain the reduction of conductance obtained for $\theta_B = 60^\circ \equiv \theta_S$ displayed in Fig.\ref{fig_sim2}a.
	
\begin{figure}[!t]
	\centering
	\includegraphics[width = 0.49\textwidth]{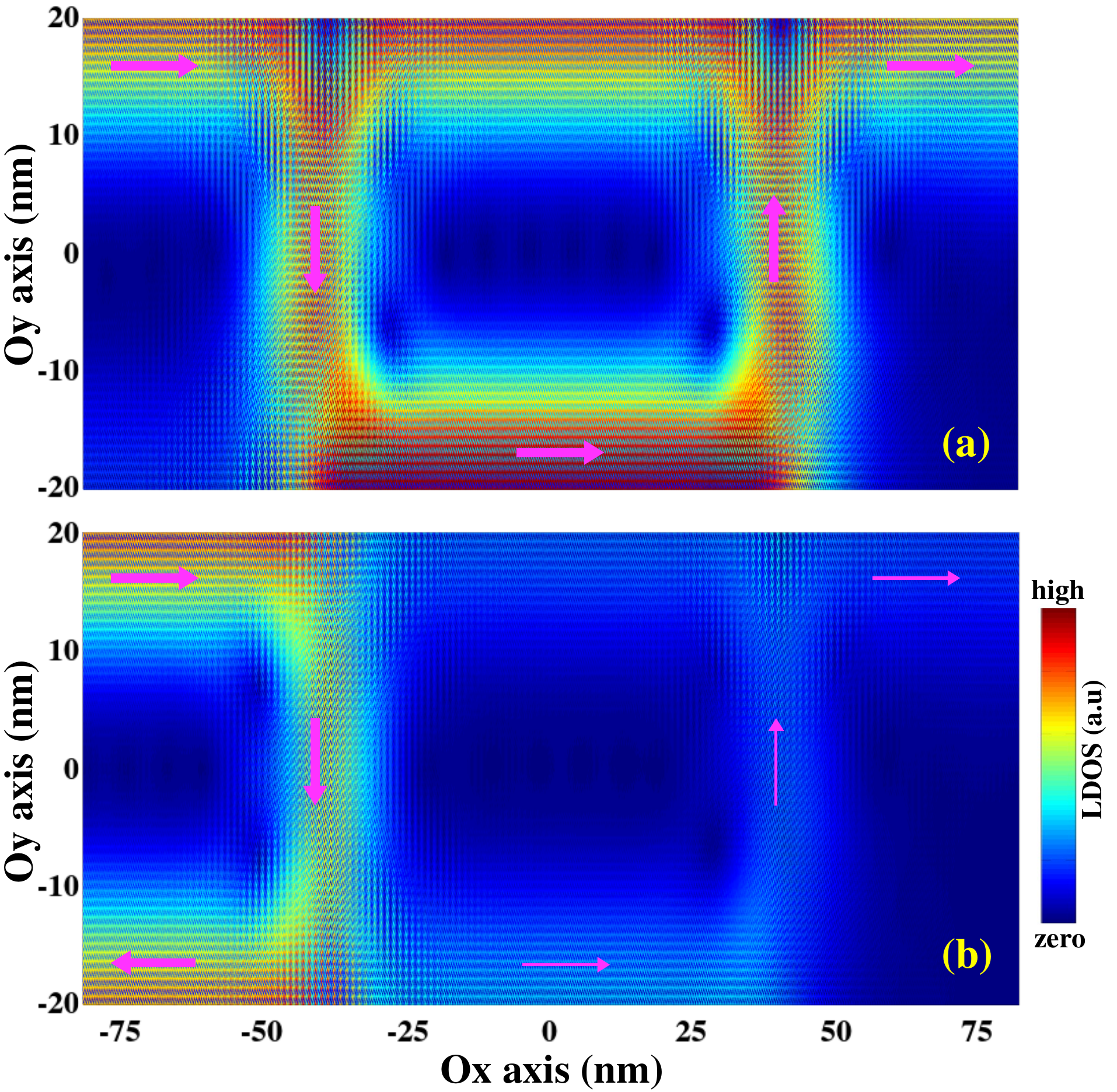}
	\caption{Left-injected local density of states at $B$ = 25.8 $T$ (a) and 27.4 $T$ (b), corresponding to conductance peak and valley, respectively (see Fig. \ref{fig_sim2}a for $\theta_B = 30^\circ$).}
	\label{fig_sim3}
\end{figure}
Most interestingly, when $0 < \theta_B < \theta_S$, two opposite $B_N$-components alternate in the terrace and facet zones, as discussed above. Similarly to the effects of $B$-field in different doped zones of graphene \textit{p-n} junctions \cite{Ozyilmaz2007,Morikawa2015,Kolasinska2016}, opposite edge states in the terrace and facet zones are created, thus inducing the strong interedge scatterings at their interface as illustrated in Fig.\ref{fig_sim2}c. As an important consequence, the conductance as a function of $B$-field represents a strong AB oscillation in the quantum Hall regime (see the case of $\theta_B = 30^\circ$ in Fig.\ref{fig_sim2}a). This result is essentially due to the interedge backscatterings diagrammatically described in Fig.\ref{fig_sim2}c and is further demonstrated by analyzing the computed left-injected local density of states in Fig.\ref{fig_sim3}, that illustrates the left-to-right electron-wave propagation. Indeed, backscatterings are almost absent (Fig.\ref{fig_sim3}a) when the phase coherence condition is satisfied, leading to conductance peaks. In the phase incoherence condition, strong interedge backscatterings (Fig.\ref{fig_sim3}b) and hence a low conductance are achieved. 

\begin{figure}[!b]
	\centering
	\includegraphics[width = 0.48\textwidth]{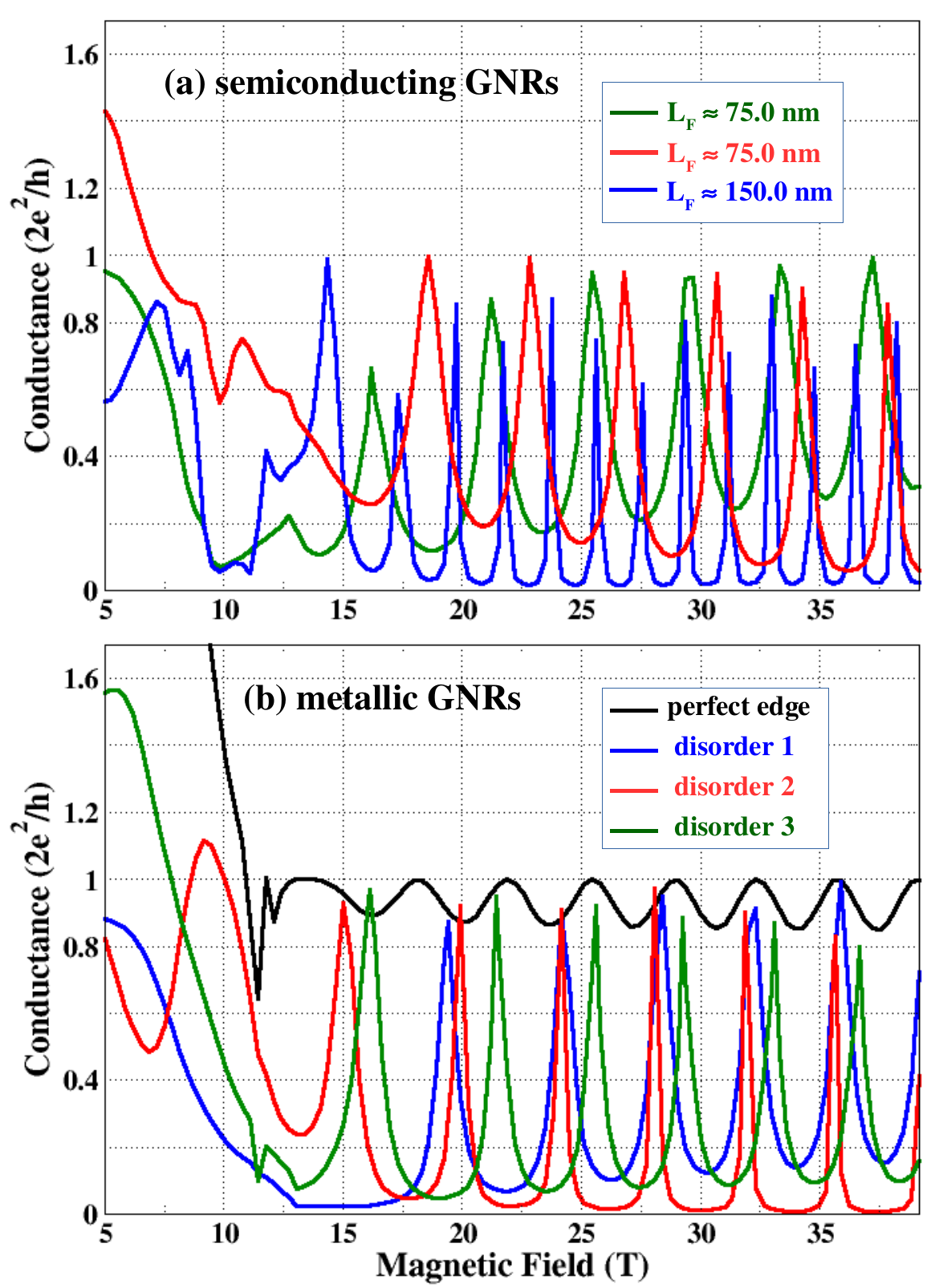}
	\caption{(a,b) Conductance as a function of $B$-field at $E_F = 75$ \textit{meV} computed for semiconducting ($N_z = 324$ \cite{Dubois2009}) and metallic ($N_z = 326$) GNR systems, respectively, with $\theta_S = 60^\circ$, $\theta_B = 30^\circ$, and $W \simeq 40$ $nm$. $L_F \simeq$ 75 \textit{nm} and 150 \textit{nm} are studied in (a) while only $L_F \simeq 75$ \textit{nm} in (b). Except for the perfect edge in (b), different disordered configurations with the variation of ribbon width $\delta$W modeled by a Gaussian auto-correlation function \cite{supplmater}, particularly, with the rsm $W_{rsm}$ = 0.6 nm and the correlation length $\xi$ = 4.8 nm are considered.}
	\label{fig_sim4}
\end{figure}

AB oscillation period observed in quantum rings with area $S$ is known to be given by $\Delta B = h/eS$ \cite{Aharonov1959}. To examine this property in the considered systems (for $\theta_B < \theta_S$), the above formula should be rewritten as  
\begin{equation}
\Delta B = \frac{h}{eS}  \frac{1}{\left| \sin(\theta_S-\theta_B) \right|}
\end{equation}
where $S$ is the area of the surface enclosed by the edge channel in the facet zone. Actually, the oscillation periods $\Delta B \simeq$ 3.61 $T$ and 1.84 $T$ are obtained in the high field regime with the facet zones of $\sim$ 3000 nm$^2$ and 6000 nm$^2$, respectively (see Fig.\ref{fig_sim4}a and additionally Figs.\ref{fig_sim2}a and \ref{fig_sim4}b). Indeed, the formula (2) predicts quite well these results of $\Delta B$ if $S \simeq$ 2332 nm$^2$ and 4640 nm$^2$ are considered, which are about 22.5 $\%$ smaller than the area of the corresponding facet zones. Note that here, the edge states are formed inside the facet zone (see Fig.\ref{fig_sim3}) and hence the value of \textit{S} in Eq.(2) to estimate $\Delta B$ is basically proportional to but smaller than the area of the facet zone. Thus, the origin of observed conductance oscillation is really the AB interference due to the interaction between edge states in both terrace and facet zones.

\begin{figure}[!h]
	\centering
	\includegraphics[width = 0.49\textwidth]{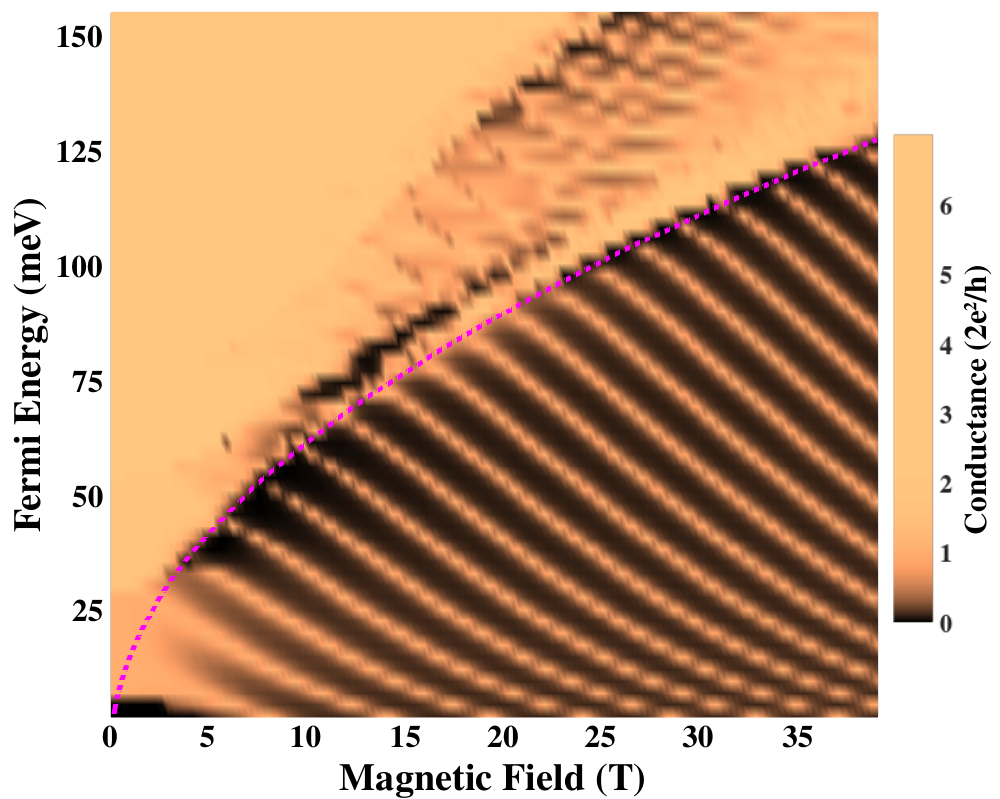}
	\caption{Conductance map with respect to $B$-field and Fermi energy obtained in the system presented in Fig.\ref{fig_sim2}a for $\theta_B = 30^\circ$. The dashed line indicates the energy level $E_1 = \min \{E_1^T, E_1^F\}$ (see text).}
	\label{fig_sim5}
\end{figure}
The observed AB interference is also found to be sensitive to some other structural parameters. First, perfect armchair GNRs can be divided in two main classes with different electronic properties, either quasi-metallic with negligible bandgap or semiconducting, depending on the number of zigzag lines $N_z$ across the ribbon width: $N_z = 3p+2$ and $N_z \neq 3p+2$ \cite{Dubois2009}, respectively. Moreover, in contrast to semiconducting ribbons, the first subband of metallic GNRs is linear, thus inducing massless fermions at low energies that contribute mainly to the transport at high \textit{B-}fields. As a consequence, a significant difference in the interference effect in the stepped systems made of these perfect GNRs is predicted.  In particular, even though the conductance oscillation is similarly observed in both cases, the effect in the metallic systems (see Fig.\ref{fig_sim4}b for the perfect edge case) is relatively weaker than the one observed in the semiconducting GNRs (see Fig.\ref{fig_sim2}a for $\theta_B = 30^\circ$).

Next, the effects of edge disorder, which are practically inevitable and known to degrade strongly the transport properties of GNRs \cite{Han2007,Han2010,Evaldsson2008,Querlioz2008,Cresti2009,Poljak2016,Fischetti2011,Djavid2014,Misawa2015,Fang2008,Goharrizi2011}, have to be evaluated. 
In this work, disorder is modeled (see the details in \cite{supplmater}) either by a Gaussian autocorrelation function (presented in Fig.\ref{fig_sim4}) or by randomly removing the edge atoms. The edge disorder indeed degrades significantly the transport at low fields. However, as shown in Figs.\ref{fig_sim4}a-b and in the Supplementary Material \cite{supplmater}, the AB oscillations at high fields are found to be much more robust under the effect of the considered disorders than the zero-field transport. This can be explained by a fascinating feature that in the quantum Hall regime, the forward and backward edge channels are spatially separated while the scatterings at these disordered edges do not allow electrons to transmit across the sample \cite{supplmater}, thus not inducing the strong backscatterings as at low fields. More interestingly, the edge disorder even eliminates the difference between the metallic and semiconducting GNR systems discussed above (i.e., comparison of the results obtained for perfect and disordered edges in Figs.\ref{fig_sim2}a and \ref{fig_sim4}). Note that a picture, similar to the metallic armchair GNR case, is also observed in zigzag GNR systems, i.e., the AB oscillation is almost invisible for perfect edges but much more pronounced in edge disordered ones \cite{supplmater}.

In Fig.\ref{fig_sim5}, the conductance as a function of both Fermi energy $E_F$ and \textit{B-}field is presented. Basically, two typical zones, $E_F \leq E_1$ and $> E_1$, are specified where $E_1 = \min \{E_1^T, E_1^F\}$ with the first Landau levels $E_1^T = \sqrt{2e\hbar v_F^2 B |\sin\theta_B|}$ and $E_1^F = \sqrt{2e\hbar v_F^2 B |\sin (\theta_S-\theta_B)|}$  \cite{Yin2017} formed in the terrace and facet zones, respectively, and $v_F$ is the Fermi velocity in graphene. In particular, strong AB oscillations are predicted for $E_F \leq E_1$ when only a single energy band is presented in both terrace and facet zones whereas the interference is blurred at higher energies. This can be explained by an inherent property of AB interference, similarly observed and demonstrated in nanoring systems \cite{Nguyen2013b,Zhang2017}, that the strong oscillations can be observed when only a single energy band contributes to the transport, otherwise the effect can be significantly disturbed by the multi-bands contribution \cite{supplmater}.

\begin{figure}[!h]
	\centering
	\includegraphics[width = 0.49\textwidth]{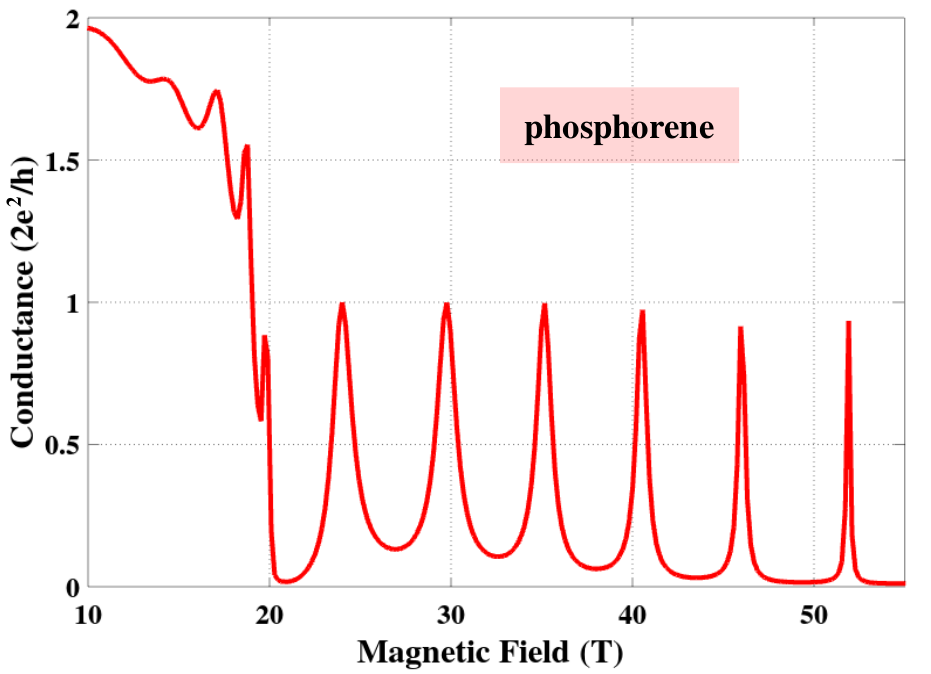}
	\caption{Conductance as a function of $B$-field obtained in a stepped phosphorene system with $\theta_S = 60^\circ$ and $\theta_B = 30^\circ$. $W \simeq 40$ \textit{nm}, $L_F \simeq 110$ \textit{nm} and the Fermi level is at 10 $meV$ above the bottom of conduction bands.}
	\label{fig_sim6}
\end{figure}
	
It is very worth noting that the interedge scatterings in the considered structures is achieved in the unipolar regime. Therefore, different from the interference phenomena reported in graphene \textit{p-n} junctions, our prediction can be achieved in both cases of semi-metallic and semiconducting materials. Indeed, a similar AB interference picture is obtained in monolayer phosphorene systems by tight-binding calculations \cite{Rudenko2014} and is presented in Fig. \ref{fig_sim6}. Moreover, structural engineering for creating stepped structures can also allow for avoiding the junction smoothness issues, that has been shown to often perturb dramatically similar quantum phenomena in graphene \textit{p-n} junctions \cite{Cheianov2006}. Finally, this predicted mechanism for achieving AB interferences, in principle, can be also applied to systems using ferromagnetic strips to create inhomogeneous \textit{B}-fields \cite{Nogaret2010}, however, obtaining sharp junctions could be a practical challenge.
	
To conclude, the magnetotransport through stepped graphene was investigated using atomistic tight-binding calculations. By applying the \textit{B-}field in a proper direction, opposite normal components of the field can be created, thus inducing opposite edge states, in the terrace and facet zones. The interedge scatterings were observed, leading to strong Aharonov-Bohm oscillations in the quantum Hall regime. The properties of this interference, depending on the carrier energy and structural parameters, were systematically clarified. Moreover, since it is observed in the unipolar regime, our prediction can be also achieved in stepped systems made of other (both semimetallic and semiconducting) 2D layered materials.
	
\textbf{Acknowledgments} - We acknowledge financial support from the F.R.S.-FNRS of Belgium through the research project (N$^\circ$ T.1077.15), from the Flag-Era JTC 2017 project "MECHANIC" (N$^\circ$ R.50.07.18.F), from the F\'{e}d\'{e}ration Wallonie-Bruxelles through the ARC on 3D nanoarchitecturing of 2D crystals (N$^\circ$ 16/21-077) and from the European Union's Horizon 2020 research and innovation program (N$^\circ$ 696656).

\newpage

\onecolumngrid

\includegraphics[page=1,scale=0.9]{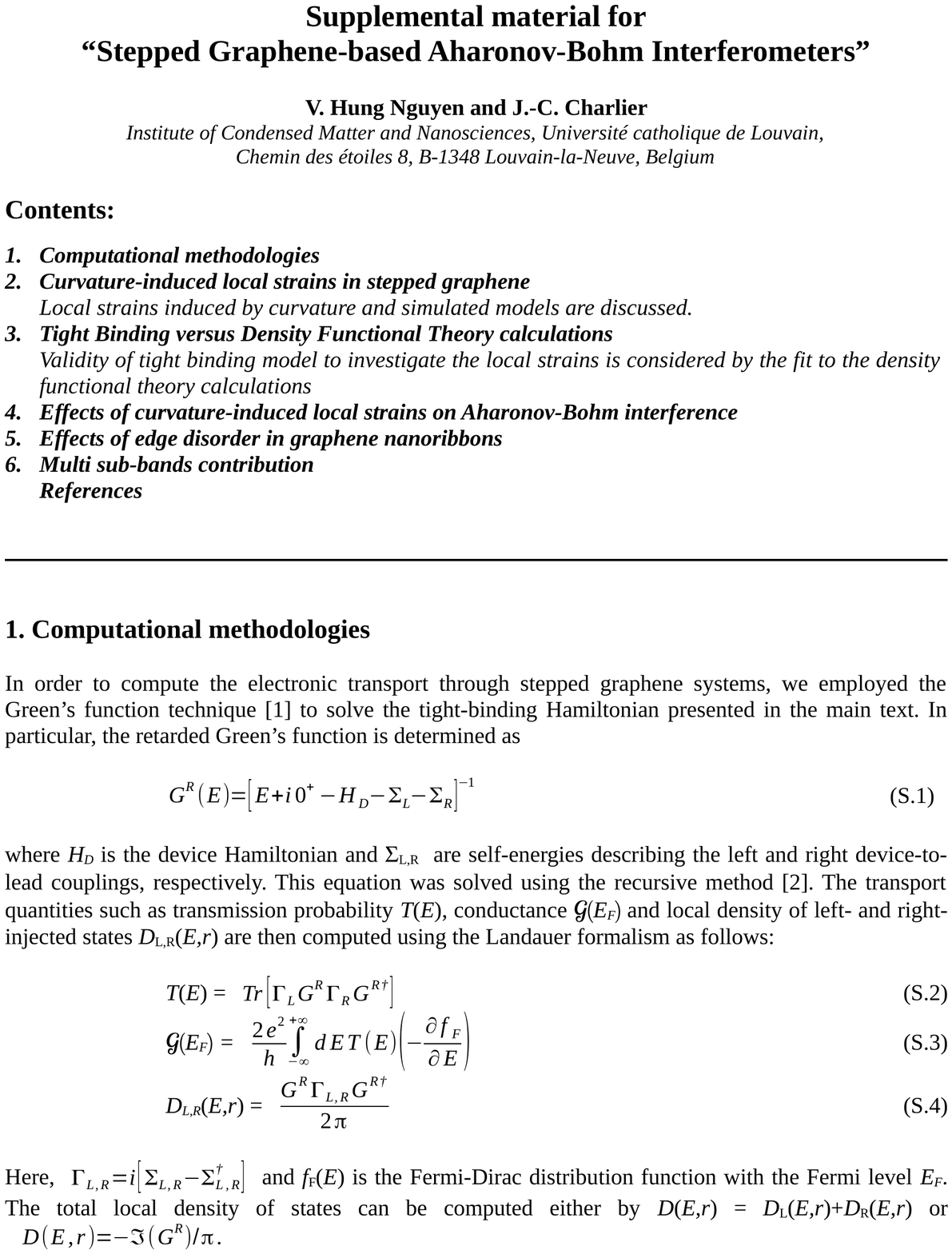} 
\includegraphics[page=2,scale=0.9]{supplmaterials.pdf} 
\includegraphics[page=3,scale=0.9]{supplmaterials.pdf} 
\includegraphics[page=4,scale=0.9]{supplmaterials.pdf} 
\includegraphics[page=5,scale=0.9]{supplmaterials.pdf} 
\includegraphics[page=6,scale=0.9]{supplmaterials.pdf} 
\includegraphics[page=7,scale=0.9]{supplmaterials.pdf} 
\includegraphics[page=8,scale=0.9]{supplmaterials.pdf} 
\includegraphics[page=9,scale=0.9]{supplmaterials.pdf} 
\includegraphics[page=10,scale=0.9]{supplmaterials.pdf} 
\includegraphics[page=11,scale=0.9]{supplmaterials.pdf} 

\end{document}